\documentclass[11pt]{article}

\usepackage[preprint]{acl}
\usepackage{xspace}
\usepackage{booktabs}
\usepackage{multirow}
\usepackage{array}

\newcommand{\ProfileDPSlot}{\textsc{DP-SLOT}\xspace}
\newcommand{\ProfileDPBlock}{\textsc{DP-BLOCK}\xspace}
\usepackage{amsmath}
\usepackage{amssymb}
\usepackage{amsthm}
\usepackage{algorithm}
\usepackage{algorithmic}
\usepackage{float}
\renewcommand{\arraystretch}{0.8}
\usepackage{makecell}
\usepackage{enumitem}
\definecolor{red}{rgb}{0.8,0,0}
\newcommand{\mypara}[1]{\noindent{\bf {#1}.}\xspace}
\newtheorem{definition}{Definition}
\newtheorem{theorem}{Theorem}
\usepackage[most]{tcolorbox}
\definecolor{GreyCool}{HTML}{E8E9E6}
\definecolor{SageRule}{HTML}{C5CDBF}
\definecolor{SageDeep}{HTML}{4E6B45}
\definecolor{BodyStrong}{HTML}{272C26}

\newcommand{\findingtagfont}{\sffamily}
\newcommand{\tracked}[2][15]{\textls[\number\numexpr#1*10\relax]{#2}}

\newtcbox{\findtag}{on line,enhanced,sharp corners,
  colback=SageDeep,colframe=SageDeep,boxrule=0pt,
  left=3.2pt,right=2pt,top=1pt,bottom=1pt,boxsep=0pt,
  fontupper=\findingtagfont\bfseries\footnotesize,coltext=GreyCool}
\newtcolorbox{findingbox}[1][Findings]{enhanced,sharp corners,
  colback=GreyCool,colframe=SageRule,boxrule=0.3pt,coltext=BodyStrong,
  overlay={\fill[SageDeep] (frame.north west) rectangle
    ([xshift=3pt]frame.south west);},
  left=10pt,right=6pt,top=5pt,bottom=5pt,
  before skip=6pt,after skip=6pt,
  before upper={\noindent\findtag{\tracked[10]{\MakeUppercase{#1}}}\hspace{0.5em}}}
\newcommand{\epstot}{\epsilon_{\mathrm{tot}}}
\definecolor{Beige}{HTML}{E4D8BE}
\definecolor{BeigeText}{HTML}{6B5A3E}
\definecolor{GreyWarm}{HTML}{ECEAE6}
\definecolor{BodyWarm}{HTML}{4F4A44}
\newtcblisting{promptblock}[1]{
  enhanced,breakable,listing only,
  title={#1},fonttitle=\bfseries\footnotesize,coltitle=BeigeText,
  title after break={#1 (continued)},
  colback=GreyWarm,colbacktitle=Beige,coltext=BodyWarm,
  colframe=Beige,boxrule=0.4pt,arc=1pt,
  left=5pt,right=5pt,top=4pt,bottom=4pt,
  before skip=6pt,after skip=6pt,
  listing options={basicstyle=\ttfamily\scriptsize,
    breaklines=true,breakatwhitespace=true,breakautoindent=false,
    columns=fullflexible,keepspaces=true,showstringspaces=false,
    numbers=none,xleftmargin=0pt,xrightmargin=0pt,frame=none,
    aboveskip=0pt,belowskip=0pt}}
\newcommand{\promptlabel}[2][12]{\needspace{#1\baselineskip}\refstepcounter{mybox}\label{#2}}
\usepackage{placeins}
\usepackage{needspace}
\usepackage{cleveref}
\newcounter{mybox}
\crefname{mybox}{Template}{Templates}
\Crefname{mybox}{Template}{Templates}

\usepackage{times}
\usepackage{latexsym}
\usepackage[T1]{fontenc}
\usepackage[utf8]{inputenc}
\usepackage{microtype}
\usepackage{inconsolata}
\usepackage{graphicx}


\title{Profile-DP: Differential Privacy for User-Controlled Profile Release in Personalized LLM Inference}

\author{Mingchen Li$^{1}$, Zifan Peng$^{2}$, Kewei Sha$^{1}$, Lavinia F Pieptea$^{1}$ \\
  {\bf Heng Fan$^{1}$, Junhua Ding$^{1}$, Meikang Qiu$^{3}$, Yunhe Feng$^{1}$} \\
  $^{1}$University of North Texas \\
  $^{2}$School of Computing, Newcastle University, UK \\
  $^{3}$Augusta University \\
  {\small\texttt{mingchenli@my.unt.edu}, \texttt{Z.Peng12@newcastle.ac.uk}} \\
  {\small\texttt{kewei.sha@unt.edu}, \texttt{laviniaflorentina.pieptea@unt.edu}} \\
  {\small\texttt{heng.fan@unt.edu}, \texttt{junhua.ding@unt.edu}} \\
  {\small\texttt{mqiu@augusta.edu}, \texttt{yunhe.feng@unt.edu}}
}

\definecolor{VarTask}{HTML}{4B5357}
\definecolor{VarNoClosure}{HTML}{C2378A}
\definecolor{VarFixedGroup}{HTML}{1F6FB2}
\definecolor{VarNoPrior}{HTML}{B8860B}
\definecolor{VarBlock}{HTML}{E6533C}
\newcommand{\mkcircle}[1]{\tikz[baseline=-.55ex]\fill[#1] (0,0) circle (.55ex);}
\newcommand{\mktriup}[1]{\tikz[baseline=-.55ex]\fill[#1] (90:.8ex)--(210:.8ex)--(330:.8ex)--cycle;}
\newcommand{\mktridown}[1]{\tikz[baseline=-.55ex]\fill[#1] (270:.8ex)--(30:.8ex)--(150:.8ex)--cycle;}
\newcommand{\mksquare}[1]{\tikz[baseline=-.55ex]\fill[#1] (-.5ex,-.5ex) rectangle (.5ex,.5ex);}
\newcommand{\mkdiamond}[1]{\tikz[baseline=-.55ex]\fill[#1] (0,.75ex)--(.6ex,0)--(0,-.75ex)--(-.6ex,0)--cycle;}

\begin{document}
\raggedbottom
\maketitle

\begin{abstract}
Personalized large language models (LLMs) increasingly rely on profile-conditioned inference, pairing public requests with rich, user-specific facts.
This paradigm exposes the user profile itself, together with the user's explicit sharing preferences.
We introduce Profile-DP, a policy-aware local differential privacy framework designed specifically for user-profile release in personalized LLM inference.
Profile-DP provably bounds the release of user-withheld facts while treating public instructions, profile schemas, and authorized facts as public side information.
By generating a privatized profile that composes with the public instruction prior to answer generation, our framework decouples formal privacy guarantees from downstream model behavior.
We develop two finite-domain mechanisms satisfying Profile-DP: \ProfileDPSlot independently perturbs protected attributes, while \ProfileDPBlock uses public semantic decisions to coordinate their release and jointly sample related attributes.
A positive, public compatibility prior connects these attributes while preserving a profile-level privacy bound.
On PEEP, \ProfileDPBlock leads the utility composite among DP methods at all 13 structured budgets and is closest to the ideal privacy--utility point in 18 of 26 format--budget settings under our composite weighting.
Direct measurements reveal complementary strengths: \ProfileDPBlock discloses less, while \ProfileDPSlot retains more authorized facts.
These results make policy-aware profile release a practical basis for studying privacy and personalization.
\end{abstract}

\section{Introduction}
LLM assistants increasingly remember who their users are.
They keep profile facts across sessions, add them to each new request, and often forward the composed prompt to a larger hosted model \citep{wang2025unveiling,li2025papillon,hartmann-etal-2024-llms}.
Users already share personal details with such systems \citep{trustnobot,DBLP:conf/iclr/Zhao0HC0D24}, yet not every stored fact belongs in every request.
Consider a user who asks an assistant to rephrase a text describing the skills they gained in their projects.
Their stored profile lists their language (French), occupation (cultural project manager), work history (internships in European cultural projects), and degree (a master's in cultural projects); they share the first two and withhold the rest.
Deleting the withheld fields is insufficient, because the shared occupation still points to the withheld work history, and LLMs readily infer such attributes from context \citep{staab2024memorizationviolatingprivacyinference}.
Rewriting the whole profile with a text-level privacy mechanism hides the withheld facts, but it also distorts the shared occupation and language that the answer depends on.

This example captures the \emph{profile-release} problem: before a prompt leaves the user's control, the system must keep the facts the user shares and protect the facts they withhold, including what related shared fields reveal about them.
Existing tools address only part of it.
PII filters and redaction target identifiers but not per-user sharing choices \citep{contextual_integrity,ghalebikesabi2024}, and language models do not reliably follow such choices on their own \citep{keepsecret,privacylens}.
Differential privacy (DP) offers formal protection \citep{dwork2006calibrating}, but DP text mechanisms define privacy over a token, span, prompt, or decoding distribution (\Cref{sec:related}), as does private in-context learning \citep{dpicl,dpicl2}.
Most of them rewrite shared and withheld facts alike; DP-Fusion \citep{thareja2025dp} follows the sharing policy but protects token spans of free text instead of profile attributes.

We introduce \textbf{Profile-DP}, a local DP notion for the user profile.
It treats every shared fact as public and formally protects only the withheld values, releasing each one as a category from a fixed public list, such as an occupation group.
Deleting a withheld value is the zero-leakage extreme; Profile-DP instead lets a coarse version of it reach the answering model with bounded leakage.
Formally, the guarantee compares profiles that agree on the instruction, the profile schema, and every shared fact, and differ only in withheld values; the adversary sees the whole prompt and may know everything the user shared.
With a sharing template, prompt construction then operates on the privatized profile through post-processing.
The guarantee therefore holds for any answering model.

\begin{figure*}[t]
\centering
\includegraphics[width=\textwidth]{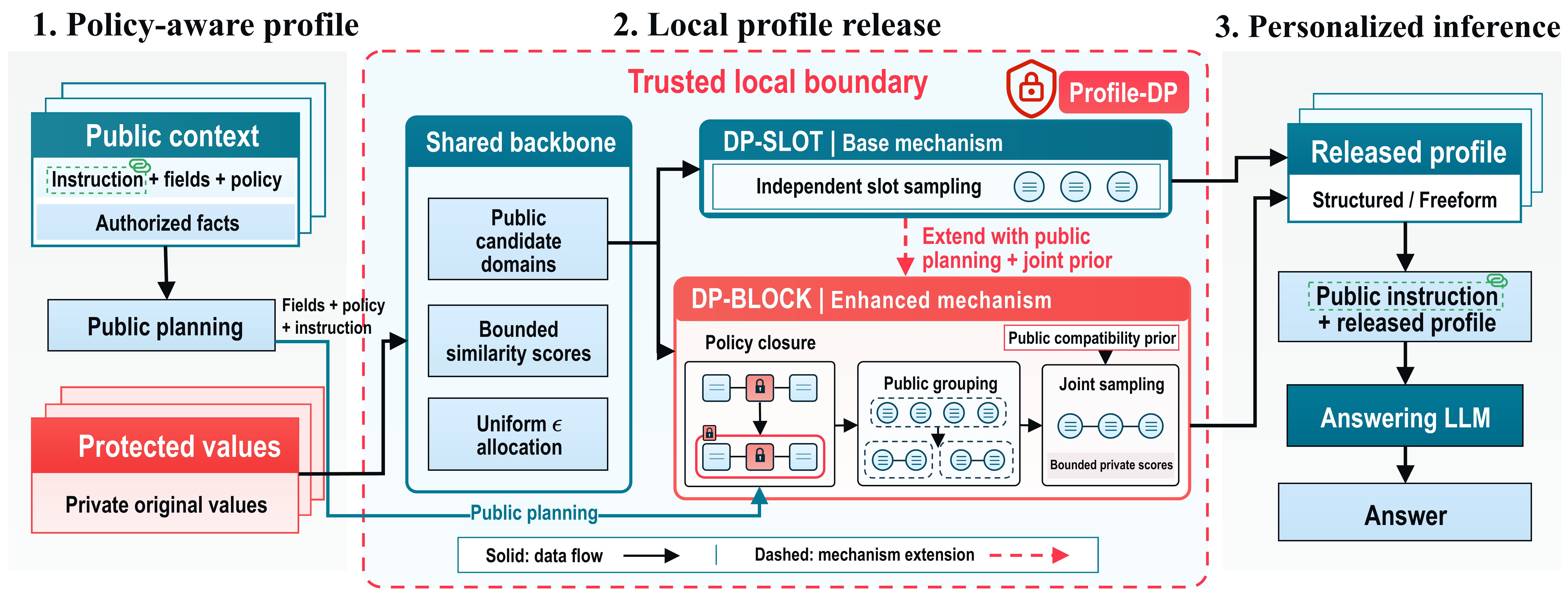}
\caption{Profile-DP overview.
Inside the trusted local boundary, a mechanism turns a policy-aware profile (public context with authorized facts, plus protected values) into a released profile; only this release and the public instruction reach the answering LLM.
Both mechanisms share a backbone of public candidate domains, bounded similarity scores, and uniform $\epsilon$ allocation; \ProfileDPBlock extends \ProfileDPSlot with public planning (policy closure and public grouping) and joint sampling with a public compatibility prior.
Solid arrows show data flow; the dashed arrow marks the mechanism extension.}
\label{fig:overview}
\vspace{-0.5cm}
\end{figure*}

We instantiate Profile-DP with two mechanisms: \ProfileDPSlot samples each withheld attribute independently, and \ProfileDPBlock adds public planning and joint sampling (\Cref{fig:overview}).
Policy closure also privatizes shared fields that would reveal a withheld one, such as the occupation above; this is an empirical safeguard beyond the formal guarantee.
Public grouping collects related withheld attributes, and joint sampling with a public compatibility prior favors plausible combinations.
Because these decisions never look at the withheld values, they consume no privacy budget, and the proof stays simple.

We evaluate on PEEP \citep{ramirez2025controlling}, which pairs real assistant requests with user profiles and synthetic per-attribute sharing preferences, against six DP text methods at 13 budget settings from 2 to 8192.
The central finding is that \ProfileDPBlock keeps what users share while bounding what they withhold.
Methods that rewrite the whole profile reach its disclosure only by discarding shared facts, keeping at most 18\% of them where \ProfileDPBlock keeps 89\%.
\ProfileDPBlock leads all DP methods in utility composite at all 13 structured budgets and is closest to the ideal privacy--utility point in 18 of 26 settings.
On freeform profiles, DP-Fusion has higher utility and is preferable up to a budget of 256, but its measured disclosure rises to 73\%, whereas that of \ProfileDPBlock stays at most 35\% at every budget.
In ablations of a task-aware variant, removing policy closure consistently increases disclosure.
The cost, relative to \ProfileDPSlot, is lower retention (89\% vs. 98\%) and slightly weaker structured answer preference.

Our contributions are:
\begin{itemize}[leftmargin=*,nosep]
\item We formalize policy-aware profile release and define Profile-DP, a guarantee over exactly the withheld values that holds for any answering model.
\item We design \ProfileDPSlot and \ProfileDPBlock, finite-domain mechanisms with a profile-level proof.
\item We show on PEEP that \ProfileDPBlock keeps most shared information with measured disclosure of at most 35\% at every budget, and identify policy closure as the most consistent disclosure control in ablations.
\end{itemize}

\section{Related Work}
\label{sec:related}
\mypara{Differentially private text sanitization}
Selective differential privacy protects policy-designated text during language-model training \citep{DBLP:conf/emnlp/ShiSCZJY22}, whereas local mechanisms sanitize inputs before downstream inference.
Metric-LDP mechanisms perturb word representations before vocabulary mapping \citep{feyisetan2020privacy,yue2021differential}, and finite-domain exponential mechanisms support text replacement \citep{carvalho2023tem,cuztext}.
LLM-facing methods rewrite prompts \citep{dpprompt}, add a second release stage to generative rewriting \citep{li2025dp}, substitute masked tokens \citep{dpmlm}, paraphrase \citep{dpparaphrase}, or mix public and private decoding distributions \citep{dpdecod}; DP-Fusion coordinates privatized distributions over policy-identified spans \citep{thareja2025dp}.
Their guarantees attach to tokens, spans, or generated text, whereas policy-labeled profiles require user authorization in the neighboring relation and a joint account of related attributes.

\mypara{User-defined privacy preferences}
Contextual integrity relates privacy to appropriate information flow \citep{contextual_integrity}.
LLM benchmarks study contextual disclosure norms \citep{keepsecret,privacylens}, information-sharing studies elicit explicit user preferences \citep{ghalebikesabi2024,cibench}, and preference-based personalization pairs requests with profile facts and sharing decisions \citep{ramirez2025controlling}.
PII redaction tools \citep{microsoft_presidio_docs} and LLM-based anonymizers \citep{staab2024large} remove identifiers without formal guarantees, although LLMs can infer personal attributes from innocuous context \citep{staab2024memorizationviolatingprivacyinference}; local mediation can protect inputs before external inference \citep{zhang-etal-2024-cogenesis,li2025papillon,hartmann-etal-2024-llms}.
These works motivate preference following but do not bound what a profile release adds about withheld values.

\mypara{Position of this work}
Profile-DP connects user-defined sharing preferences to local differential privacy \citep{duchi2013local}: it treats authorized facts as public side information, protects withheld values through a profile-level neighboring relation, and coordinates related attributes through public decisions before joint sampling, in the spirit of joint exponential mechanisms \citep{gillenwater2022jointEM}.
Our evaluation follows the same boundary, measuring protected disclosure alongside authorized retention and downstream utility.

\section{Preliminaries and Problem Setting}
\label{sec:prelim}
\mypara{Differential privacy}
A randomized mechanism $\mathcal M$ is $\epsilon$-differentially private if, for neighboring inputs $u\simeq u'$ and every measurable event $S$,
\begin{equation*}
 \Pr[\mathcal M(u)\in S]\leq e^\epsilon\Pr[\mathcal M(u')\in S].
\end{equation*}
The exponential mechanism samples $y$ from a fixed support with probability proportional to $\exp(\epsilon s(u,y)/(2\Delta s))$, where $\Delta s$ bounds the score change between neighbors \citep{mcsherry2007em}.
Sequential releases compose additively, and post-processing preserves the guarantee.

\mypara{Policy-aware profiles}
A profile is $u=\{x_i=(e_i,f_i,v_i,a_i)\}_{i=1}^{n}$ with $n$ attributes.
Each attribute has an entity $e_i$, field $f_i$, value $v_i$, and release preference $a_i\in\{\text{shared},\text{withheld}\}$.
Let $P=\{i:a_i=\text{withheld}\}$ be the withheld (protected) set; shared values are authorized for release.
The public instruction $q$ is the user's task request.
We collect the public context as
\begin{equation*}
 c=\left(q,\{(e_i,f_i,a_i)\}_{i=1}^{n},\{v_i:i\notin P\}\right).
\end{equation*}
Two profiles satisfy $u\simeq_c u'$ when they share $c$; their withheld values may differ arbitrarily.

\begin{definition}[Profile-DP]
For each fixed public context $c$, a local profile-release mechanism $\mathcal M$ satisfies $\epsilon$-Profile-DP if, for every $u\simeq_c u'$ and output event $S$,
\begin{equation*}
 \Pr[\mathcal M(u;c)\in S]\leq e^\epsilon\Pr[\mathcal M(u';c)\in S].
\end{equation*}
\end{definition}
The adversary observes $c$ and the released profile, and may also observe downstream responses.
The guarantee bounds the additional evidence supplied by the release about withheld values, conditional on $c$.
Information inferable from the instruction or shared facts is part of that conditioning.
The local mechanism accesses the original profile inside the user's trusted boundary.

\mypara{Local profile release}
The mechanism produces a privatized profile $\tilde u$.
A deterministic prompt template $R$ renders it either as a structured profile or as freeform text and combines it with the instruction:
\begin{equation*}
 p=\operatorname{Compose}(q,R(\tilde u)),\quad o=F(p),
\end{equation*}
where $F$ is the answering model and $o$ its response.
\Cref{app:algorithm} gives the complete release algorithm.

\section{Method}
\label{sec:method}
Both mechanisms build on a shared backbone of public candidate domains, bounded similarity scores, and uniform budget allocation (\Cref{fig:overview}).
\ProfileDPSlot samples each withheld attribute independently, and \ProfileDPBlock extends it with public planning and joint sampling.

\subsection{Shared Backbone}
\label{sec:domains-scores}
\mypara{Public candidate domains}
Finite public domains give neighboring profiles the same output support, and public taxonomies supply interpretable categories.
For field $f$, we fix a candidate set $\mathcal C_f$ from public classification systems, such as UN geographic regions \citep{un_stats_m49} and ISCO occupation groups \citep{ilo_isco08}.
A withheld attribute is \emph{supported} when its field has such a domain; unsupported withheld values and protected identifiers receive a fixed suppression symbol.
\Cref{app:domains} lists the domains and their sources.

\mypara{Bounded similarity scores}
To retain useful meaning, we score each public candidate against the original value.
Let $E(t)$ be the unit-normalized embedding of text $t$, and $\mathcal T_f(y)$ the official prototype texts for candidate $y$, or its label when no prototype text is available.
We compute
\begin{align*}
 \kappa_f(v,y)&=\max_{t\in\mathcal T_f(y)}\langle E(v),E(t)\rangle,\\
 s_f(v,y)&=\operatorname{clip}_{[0,1]}((1+\kappa_f(v,y))/2).
\end{align*}
Because $s_f\in[0,1]$, its global sensitivity is at most one.
For example, a cultural project manager can receive a high score for the public category ``Managers,'' while remaining eligible for every other occupational category.
The original value is compared directly with candidate prototypes, preserving semantic distinctions within a shared public support.
\Cref{app:score} specifies the embedding model and special-case scores.

\mypara{Uniform allocation}
Let $I\subseteq P$ contain the supported withheld attributes.
A total budget $\epsilon$ is allocated uniformly as $\epsilon_i=\epsilon/|I|$ for $i\in I$, so the composition bound is $\sum_{i\in I}\epsilon_i\leq\epsilon$.

\subsection{\ProfileDPSlot: Independent Release}
Independent release preserves shared facts while randomizing withheld values; each withheld attribute forms one \emph{slot}.
For each supported slot $i\in I$, \ProfileDPSlot samples a public candidate with the exponential mechanism:
\begin{equation*}
 \Pr[\tilde v_i=y]
 =\frac{\exp(\epsilon_i s_{f_i}(v_i,y)/2)}
 {\sum_{z\in\mathcal C_{f_i}}\exp(\epsilon_i s_{f_i}(v_i,z)/2)}.
\end{equation*}
The mechanism favors candidates close to the private value while leaving every public candidate possible, so an occupation is released as a broad occupational group, for example.
Shared values are copied, and unsupported withheld values are suppressed; if $I$ is empty, release consists entirely of these deterministic operations.
\ProfileDPSlot thus separates authorization from randomization: shared values pass through unchanged, and every supported withheld attribute receives the same budget.

\subsection{\ProfileDPBlock: Joint Release}
\label{sec:block}
Independent sampling treats each slot in isolation.
Two difficulties arise at the profile level: a shared field can reveal information about a withheld field, and independently sampled fields can describe an unlikely combination.
\ProfileDPBlock addresses them with public planning, namely policy closure and public grouping, followed by joint sampling with a public compatibility prior.
All three decisions use a small binary-decision model that returns yes/no probabilities with low latency (Laya; \Cref{app:planning,app:thresholds}); none of them sees a withheld value.
The decision model sees public field descriptions for closure and grouping, and public candidate values for compatibility.

\mypara{Policy closure}
A withheld work history can also be revealed by a shared occupation (\Cref{tab:release-example}).
Closure subjects such related sources to randomization or suppression before release.
For fields belonging to the same entity, the decision model scores whether revealing a source field would strongly constrain a withheld target, using the instruction, field identities, and preference labels.
Starting from $P$, sources with scores at least $\tau_C$ join the privatization set, and the process continues to a fixed point $P^+$.
Closure preserves every original withheld label and can additionally withhold a shared field that describes the same information.
The formal neighboring relation still varies only the original withheld values; original shared values remain fixed in $c$.
Closure thus decides which related facts are released, while the formal guarantee still concerns only the user's original withheld values.

\mypara{Public grouping}
Joint selection is useful when fields describe related aspects of a person.
For example, occupation and work history can carry overlapping professional information, motivating a shared release decision.
Let $I^+\subseteq P^+$ contain the supported attributes after closure, with uniform budgets $\epsilon_i=\epsilon/|I^+|$.
A second decision scores the benefit of selecting two fields together, and edges with scores above $\tau_G$ form connected groups in $I^+$; oversized groups are split into small blocks (\Cref{app:algorithm}).
Each block $b$ uses the full Cartesian support $\mathcal C_b=\prod_{i\in b}\mathcal C_{f_i}$.
Grouping determines where dependence can be modeled; the compatibility prior determines how it affects sampling.

\mypara{Joint sampling with a compatibility prior}
Grouping becomes an active joint mechanism through a public compatibility score $J_b(y_b)\in[0,1]$, which averages the decision model's plausibility scores for candidate pairs along retained edges.
For example, related occupation and work fields can favor compatible professional categories when sampled together.
The corresponding positive prior $\pi_b(y_b)\propto\exp(\gamma J_b(y_b))$ serves as a public base measure for the exponential mechanism:
\begin{align*}
 L_b(u,y_b)&=\gamma J_b(y_b)
       +\frac12\sum_{i\in b}\epsilon_i s_{f_i}(v_i,y_i),\\
 \Pr[\tilde v_b=y_b]&=
       \frac{\exp L_b(u,y_b)}
       {\sum_{z_b\in\mathcal C_b}\exp L_b(u,z_b)}.
\end{align*}
The prior favors compatible tuples while every candidate tuple retains positive probability, including unusual combinations.
For constant $J_b$, the distribution factorizes into independent \ProfileDPSlot draws with the same effective privatization set and attribute budgets, so the compatibility prior supplies the joint dependence.
Private values enter solely through $s_{f_i}$, preserving the bounded influence of each similarity score.
We use $\tau_C=0.55$, $\tau_G=0.35$, and $\gamma=2$.
\Cref{tab:release-example} shows a real release in which closure protects a shared field that reveals a withheld one.

\begin{table}[t]
\centering\scriptsize
\caption{A real release at $\epstot=64$.
Both mechanisms release each withheld value here as a sampled public category, which need not match the original.
Closure ($\dagger$) also privatizes the shared occupation, which reveals the withheld work history; \ProfileDPSlot copies it (\textcolor{red!75!black}{red}).}
\label{tab:release-example}
\setlength{\tabcolsep}{2.5pt}
\renewcommand{\arraystretch}{1.05}
\begin{tabular}{@{}c@{\hspace{3pt}}l>{\raggedright\arraybackslash}p{.25\columnwidth}>{\raggedright\arraybackslash}p{.23\columnwidth}>{\raggedright\arraybackslash}p{.21\columnwidth}@{}}
\toprule
& Field & Original & \ProfileDPSlot & \ProfileDPBlock\\
\midrule
\multirow{3}{*}{\rotatebox[origin=c]{90}{\textit{Shared}}}
& Languages & French & French & French\\
& Occupation$^\dagger$ & Cultural Project Manager & \textcolor{red!75!black}{Cultural Project Manager} & Managers\\
\midrule
\multirow{5}{*}{\rotatebox[origin=c]{90}{\textit{Withheld}}}
& Work & Internships in European cultural projects & Plant and machine operators & Clerical support workers\\
& Education & Master's degree in Cultural Projects & ISCED 4 (post-secondary) & ISCED 6 (bachelor's)\\
\bottomrule
\end{tabular}
\vspace{-0.4cm}
\end{table}

\subsection{Privacy Guarantee and Reuse}
\begin{theorem}[Profile-release guarantee]
\label{thm:release}
Fix public context $c$, public supports, and a public planning rule.
If all private scores lie in $[0,1]$ and $\sum_i\epsilon_i\leq\epsilon$, then \ProfileDPSlot and \ProfileDPBlock with fresh internal randomness satisfy pure $\epsilon$-Profile-DP.
\end{theorem}
For a block, changing withheld values alters each log weight by at most half the sum of their allocated budgets.
The normalizer contributes the same bound.
Composition across blocks gives the profile guarantee; \Cref{app:proof} provides the full proof.

Prompt construction and answer generation are post-processing of the release and public context.
Reusing a single release across public requests retains its original bound, provided subsequent computation accesses that release and the fixed public information.
Generating $T$ fresh releases of the same withheld values instead incurs a bound of $\sum_{t=1}^{T}\epsilon^{(t)}$.
Uniform allocation makes a request-independent \ProfileDPSlot release reusable across tasks; the public planning of \ProfileDPBlock can depend on the request.

\begin{figure*}[t]
\centering
\includegraphics[width=\textwidth]{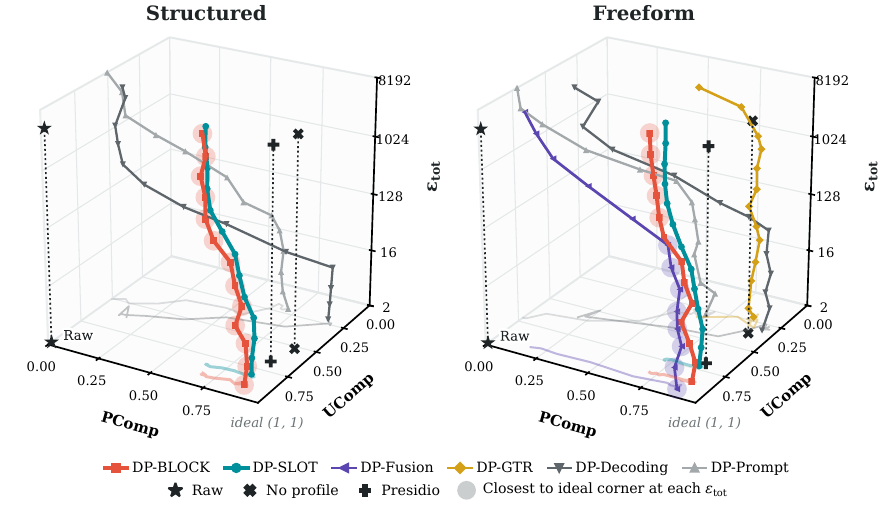}
\caption{Privacy--utility trajectories across settings of the total-budget parameter $\epstot$ (\Cref{sec:experiments}) for structured (left) and freeform (right) profiles.
Each curve places a method at its privacy (PComp) and utility (UComp) composites, with $\epstot$ on the vertical axis; the floor shows the projected trade-off, and the corner nearest the reader is the ideal point $(1,1)$.
Halos mark the DP method closest to that point at each budget, and vertical poles mark budget-independent controls.
DP-MLM and DP-Paraphrase, whose UComp stays below $0.35$, are omitted for readability but remain in the normalization and the closest-point comparison.
\ProfileDPBlock is closest at every structured budget; for freeform profiles, DP-Fusion is closest up to $\epstot=256$ and \ProfileDPBlock beyond.}
\label{fig:composites}
\vspace{-0.4cm}
\end{figure*}

\section{Experiments}
\label{sec:experiments}
\subsection{Experimental Setup}
\mypara{Data and task}
We evaluate a fixed random sample of 500 PEEP records, whose user requests and profiles derive from WildChat \citep{ramirez2025controlling,DBLP:conf/iclr/Zhao0HC0D24}, with PEEP's synthetic sharing preferences.
An LLM preprocessing step extracts a task-only public instruction from each request, and the extracted values and preference labels form the profile (\Cref{app:data}).
Every method releases either a \textbf{structured} profile or a \textbf{freeform} rendering of the same attributes, and all methods receive the same records and instruction template.

\mypara{Compared methods}
We compare \ProfileDPSlot and \ProfileDPBlock with DP-Prompt \citep{dpprompt}, DP-Decoding \citep{dpdecod}, DP-MLM \citep{dpmlm}, and DP-Paraphrase \citep{dpparaphrase}, and, on freeform profiles, with DP-GTR \citep{li2025dp} and DP-Fusion \citep{thareja2025dp}.
Structured baselines perturb individual values while preserving profile keys, and freeform baselines process the profile text.
DP-Fusion protects policy-identified withheld spans, whereas the other baselines rewrite shared and withheld text alike.
Controls include the original profile (Raw), the instruction alone (No profile), and rule- and NER-based PII redaction with Presidio \citep{microsoft_presidio_docs}; \Cref{app:baselines} adds two further suppression controls and gives every release policy.

\mypara{Budgets and models}
We evaluate 13 settings of the total-budget parameter $\epstot$, doubling from $2$ to $8192$.
For our methods, $\epstot=\epsilon$ is the total profile budget divided among withheld attributes; baselines use the same grid with their native allocation and composition rules (\Cref{app:budgets}).
The lower range probes stronger randomization, and the upper range examines useful-release behavior under progressively weaker formal bounds.
Qwen3.6-27B-FP8 generates responses \citep{qwen3.6-27b}, and Qwen3.6-35B-A3B-FP8 judges disclosure and response quality \citep{qwen36_35b_a3b}, both with deterministic decoding (\Cref{app:models}).
Results use one profile-sampling seed; the interval analysis resamples evaluation records.

\mypara{Metrics}
Our primary direct diagnostics are \textbf{Protected Profile Disclosure Rate (PPDR)} and \textbf{Raw-relative Utility Loss (RUL)}, both lower-is-better.
PPDR measures the fraction of withheld attributes revealed exactly or semantically in a released profile.
RUL compares the generated answer with an answer produced from the original profile, averaging judgments in both presentation orders; it ranges from $-1$ (method preferred) to $1$ (raw preferred).
Profile retention (PR) measures the fraction of shared values preserved in the released profile.
Protected-attribute recovery (PARR), answer disclosure (AD), answer retention (AR), and reference-conditioned perplexity (RCP), which measures support for the dataset reference answer, complete the evaluation (\Cref{app:metrics,app:templates}).

To summarize the complementary objectives, \textbf{Privacy Composite (PComp)} averages normalized PPDR, PARR, and AD.
\textbf{Utility Composite (UComp)} weights the win-or-tie rate against Raw answers (PW) by $2/5$ and RCP, PR, and AR by $1/5$ each.
Both are higher-is-better; we report them jointly and identify, at each budget, the DP method closest to the ideal point $(\mathrm{PComp},\mathrm{UComp})=(1,1)$.
Normalization uses one range per format over the fixed DP comparison set and all 13 budgets, so changes across budgets reflect each method's own measurements; direct metrics retain their original scales.
Confidence intervals and sensitivity analyses appear in \Cref{app:ci,app:robustness}.

\begin{figure*}[t]
\centering
\includegraphics[width=\textwidth]{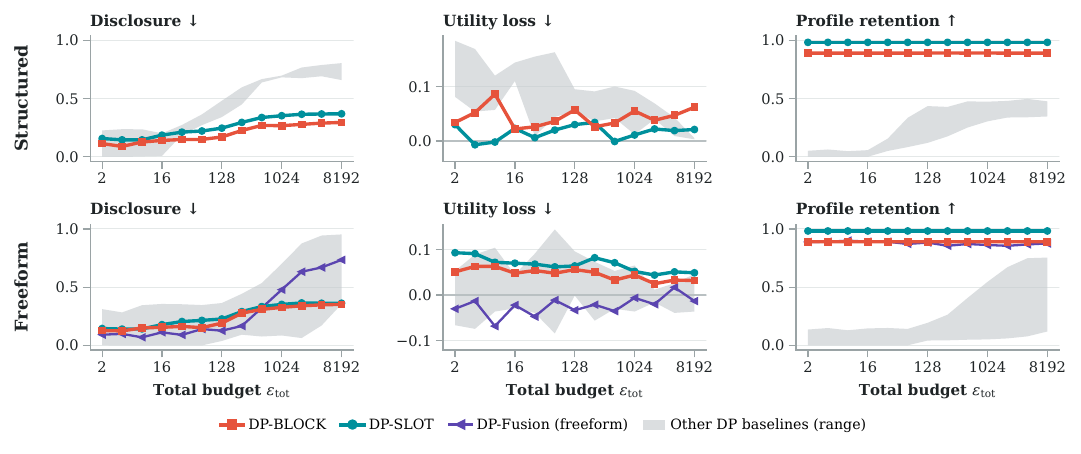}
\caption{Direct diagnostics across budgets.
Lower is better for protected-profile disclosure (PPDR) and raw-relative utility loss (RUL); higher is better for profile retention of shared values (PR).
Rows show structured and freeform profiles.
The gray band spans the other displayed DP baselines (DP-Prompt, DP-Decoding, and, for freeform profiles, DP-GTR); attribute recovery appears in \Cref{fig:additional}, and budget-independent controls appear in \Cref{fig:composites} and \Cref{tab:utility-context}.
\ProfileDPBlock keeps PPDR at most $0.35$ and $89\%$ of shared values at every budget.}
\label{fig:diagnostics}
\vspace{-0.4cm}
\end{figure*}

\subsection{Privacy--Utility Trade-off}
The composite view tests whether a released profile retains useful information while reducing exposure.
In \Cref{fig:composites}, \ProfileDPBlock leads UComp among the evaluated DP methods at every structured budget and is closest to the ideal point at all 13 structured budgets, including the five budgets from $2$ to $32$, where private similarity scores exert less influence on sampling.
For freeform profiles, DP-Fusion leads UComp throughout and is closest to the ideal point up to $\epstot=256$, with \ProfileDPBlock second at each low budget; \ProfileDPBlock is closest from $512$ onward.
These counts are unchanged when the recovery probe is removed from PComp (\Cref{app:robustness}).
The two methods differ in how disclosure grows with the budget: DP-Fusion's PPDR rises from $0.09$ to $0.73$ and exceeds that of \ProfileDPBlock from $\epstot=512$ (\Cref{fig:diagnostics}), whereas \ProfileDPBlock stays between $0.12$ and $0.35$ at every budget.
DP-Fusion's advantage is therefore concentrated at small budgets, while \ProfileDPBlock keeps disclosure stable across the full budget range.
These point estimates use the declared weights, counting ties as answer wins.
\begin{findingbox}
\ProfileDPBlock offers the strongest structured utility composite, including at low budgets, and a competitive privacy--utility balance across formats under our weighting.
DP-Fusion is preferable for freeform profiles at small budgets.
\end{findingbox}

\begin{figure}[!tp]
\centering
\begin{minipage}{\columnwidth}
\centering\footnotesize
\captionof{table}{Component ablations on the ablation set.
\checkmark{} enables policy closure (Clo.), public grouping (Grp.), the public compatibility prior (Cmp.), or uniform allocation (Uni.).
Protected-profile disclosure (PPDR) and raw-relative utility loss (RUL) are five-budget means, both lower is better; \Cref{app:components} gives per-budget results.}
\label{tab:component-main}
\setlength{\tabcolsep}{2.4pt}
\renewcommand{\arraystretch}{1.08}
\resizebox{\columnwidth}{!}{%
\begin{tabular}{lcccccccc}
\toprule
& \multicolumn{4}{c}{Components} & \multicolumn{2}{c}{Structured} & \multicolumn{2}{c}{Freeform}\\
\cmidrule(lr){2-5}\cmidrule(lr){6-7}\cmidrule(lr){8-9}
Variant & Clo. & Grp. & Cmp. & Uni. & PPDR & RUL & PPDR & RUL\\
\midrule
\mkcircle{VarTask}~Task & $\checkmark$ & $\checkmark$ & $\checkmark$ & $\times$ & .153 & .049 & .202 & .009\\
\mktriup{VarNoClosure}~No-Closure & $\times$ & $\checkmark$ & $\checkmark$ & $\times$ & .204 & .037 & .258 & $-$.020\\
\mkdiamond{VarFixedGroup}~Fixed-Group & $\checkmark$ & $\times$ & $\checkmark$ & $\times$ & .158 & .018 & .191 & $-$.014\\
\mktridown{VarNoPrior}~No-Prior & $\checkmark$ & $\checkmark$ & $\times$ & $\times$ & .176 & .016 & .213 & .014\\
\midrule
\mksquare{VarBlock}~Block & $\checkmark$ & $\checkmark$ & $\checkmark$ & $\checkmark$ & .174 & .004 & .205 & .000\\
\phantom{\mksquare{VarBlock}}~Slot & $\times$ & $\times$ & $\times$ & $\checkmark$ & .196 & .016 & .260 & $-$.017\\
\bottomrule
\end{tabular}}
\end{minipage}

\vspace{3mm}
\includegraphics[width=\columnwidth]{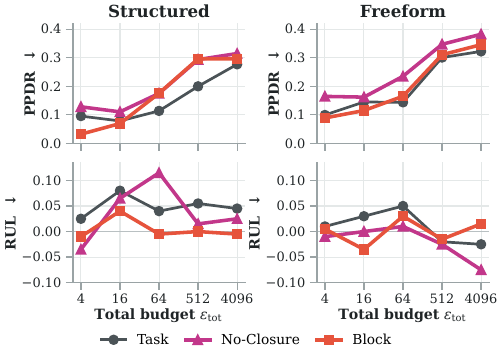}
\caption{Key ablations on the ablation set at all five tested budgets.
Rows show protected-profile disclosure (PPDR; 64 eligible records) and raw-relative utility loss (RUL; 100 records), both lower is better; markers follow \Cref{tab:component-main}.
Task has lower PPDR than No-Closure at every budget in both formats.}
\label{fig:component-main}
\end{figure}

\subsection{Privacy Diagnostics}
The composite ranking combines two questions that we now examine directly: what each release discloses about withheld values, and what it preserves of shared ones.
Disclosure measures detected withheld facts; recovery probes a fixed attacker supplied with public context.
In \Cref{fig:diagnostics}, \ProfileDPBlock has lower PPDR than \ProfileDPSlot at every structured budget and most freeform budgets, and across all budgets its PPDR stays at most $0.29$ for structured and $0.35$ for freeform profiles.
Recovery remains substantial even when measured disclosure is low, reflecting the information available to the fixed attacker through public context (\Cref{fig:additional}).
Suppression controls yield lower disclosure alongside lower authorized retention (\Cref{tab:utility-context}).
\begin{findingbox}
\ProfileDPBlock reduces measured profile disclosure relative to \ProfileDPSlot, while stronger filtering achieves further reductions at the cost of retained information.
\end{findingbox}

\subsection{Utility Preservation}
\begin{figure}[t]
\centering
\includegraphics[width=\columnwidth]{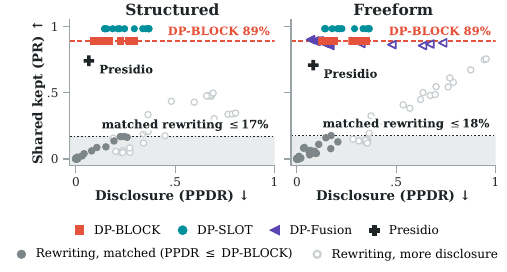}
\caption{Protected-profile disclosure (PPDR) versus profile retention of shared values (PR) per method and budget (upper left is better).
Filled gray points are rewriting baselines at budgets where they disclose no more than \ProfileDPBlock; hollow points are budgets where they disclose more.
At matched disclosure, rewriting keeps at most 18\% of shared values verbatim (dotted), versus 89\% for \ProfileDPBlock (dashed).}
\label{fig:matched-retention}
\vspace{-0.4cm}
\end{figure}

Lower disclosure matters only if shared information survives, so we separate answer quality from retention.
\Cref{fig:diagnostics} shows higher profile retention for \ProfileDPSlot than for \ProfileDPBlock, whose closure can withhold related shared fields.
\Cref{fig:matched-retention} compares retention at matched disclosure: whenever a whole-profile rewriting method discloses no more than \ProfileDPBlock at the same budget, it keeps at most $18\%$ of shared values, whereas \ProfileDPBlock keeps $89\%$.
DP-Fusion, which follows the sharing policy, matches both up to a budget of 256, consistent with its small-budget advantage in \Cref{fig:composites}.
Presidio redaction discloses less but keeps fewer shared values and offers no formal guarantee (\Cref{fig:matched-retention}).
\noindent Across the full-cohort curves, \ProfileDPBlock has lower freeform RUL throughout, while \ProfileDPSlot generally has lower structured RUL.
\Cref{tab:utility-context} lists these values at $\epstot=64$.
\Cref{app:robustness} decomposes the answer preferences.
\begin{findingbox}
Methods that rewrite the whole profile match the disclosure of \ProfileDPBlock only by discarding shared values.
\ProfileDPSlot favors shared-information retention, while the two complete systems have complementary response-utility strengths across formats.
\end{findingbox}

\subsection{Component Analysis}
Because \ProfileDPBlock differs from \ProfileDPSlot only through public planning and joint sampling, we isolate policy closure, public grouping, the compatibility prior, and allocation.
We fix a 100-record subset as the ablation set and evaluate every variant in \Cref{tab:component-main} on it.
Task allocates the budget by each withheld field's scored relevance to the instruction and serves only as the ablation reference; the evaluated \ProfileDPBlock uses uniform allocation.
\Cref{fig:component-main} plots the key ablations against Task.
Task has lower PPDR than No-Closure in both formats at every plotted budget; this intervention includes changes to the privatization set and budget allocation.
Averaged over budgets, removing closure raises PPDR from $0.153$ to $0.204$ for structured profiles and from $0.202$ to $0.258$ for freeform profiles.
Fixed-Group and No-Prior trace different disclosure--utility operating points (\Cref{fig:components}), and uniform Block remains competitive with Task.
\Cref{app:components} gives the complete ablation, including a matched-randomness prior diagnostic and full-cohort allocation comparisons.
\begin{findingbox}
Closure is the most consistent disclosure-control component in the task-aware comparison.
The grouping and allocation alternatives offer budget-dependent privacy--utility trade-offs.
\end{findingbox}

\section{Conclusion}
Profile-DP makes user-controlled profile release an explicit privacy boundary in personalized LLM inference.
\ProfileDPSlot and \ProfileDPBlock instantiate this boundary with independent and jointly sampled exponential mechanisms, whose public planning retains a transparent profile-level proof.
On PEEP, \ProfileDPBlock leads the utility composite among DP methods on structured profiles and is closest to the ideal privacy--utility point in 18 of 26 settings under the declared weights.
It keeps 89\% of shared values, versus at most 18\% for whole-profile rewriting at matched disclosure, while its disclosure stays at most 35\% at every budget; ablations identify policy closure as its most consistent disclosure control.
The trade-offs remain explicit: \ProfileDPSlot retains more shared facts, and DP-Fusion is preferable for freeform profiles at small budgets.
The formulation and evidence provide a foundation for policy-aware personalization with explicit release guarantees.

\clearpage
\section{Limitations}
\mypara{Candidate coverage and semantic proximity}
Release quality depends on the coverage and granularity of the public domains.
Small categorical supports can leave substantial semantic information even after randomization.
Profile-DP bounds only the evidence a release adds to the public context.
Policy closure privatizes shared fields that strongly constrain a withheld one, but weaker associations in shared facts and the instruction can still support inference of withheld attributes.

\mypara{Evaluation scope}
The experiments cover one policy-aware benchmark with a fixed local answer model and judge.
Answer judgments measure task quality, and retention measures the availability of shared information; correct use of personalized facts requires a separate, profile-informed assessment.
The composite weights encode one operational preference among these measurements.
Uniform allocation was chosen after inspecting the component results, so the comparisons describe operating points observed on this benchmark.

\mypara{Privacy calibration and repeated use}
Large total budgets give weak formal bounds, and their privacy is assessed through empirical disclosure measurements.
Baselines keep their native adjacency and calibration, so the same $\epstot$ does not imply equal formal guarantees across methods.
Fresh releases accumulate privacy loss, and because public planning can depend on the request, a \ProfileDPBlock release is not automatically reusable across tasks.
Results use one profile-sampling seed, so seed-to-seed variation of the mechanisms is not quantified.

\bibliography{custom}
\appendix
\setcounter{dbltopnumber}{3}\setcounter{topnumber}{3}\setcounter{totalnumber}{5}
\renewcommand{\dbltopfraction}{0.9}\renewcommand{\topfraction}{0.9}
\renewcommand{\textfraction}{0.07}
\renewcommand{\dblfloatpagefraction}{0.75}\renewcommand{\floatpagefraction}{0.75}
\section{Profile-DP Proof and Composition}
\label{app:proof}
\mypara{Fixed public decisions}
Fix $c$ and $u\simeq_c u'$.
The closure, privatization set, public domains, partition, and budgets are identical for both inputs.
Compatibility scores are also identical because their inputs are public candidate tuples.
Any originally shared attributes promoted by closure have equal original values under the neighboring relation.

\mypara{Block likelihood ratio}
Let $A_b=\{i\in b:v_i\neq v_i'\}$ and $h_b=\frac12\sum_{i\in A_b}\epsilon_i$.
The bounded score gives
\begin{equation*}
 |L_b(u,y_b)-L_b(u',y_b)|\leq h_b.
\end{equation*}
Writing $Z_b(u)=\sum_{z_b\in\mathcal C_b}\exp L_b(u,z_b)$, the same pointwise bound implies
\begin{equation*}
 e^{-h_b}Z_b(u')\leq Z_b(u)\leq e^{h_b}Z_b(u').
\end{equation*}
Every tuple has positive probability, and hence
\begin{align*}
 \frac{\Pr[\tilde v_b=y_b\mid u,c]}
      {\Pr[\tilde v_b=y_b\mid u',c]}
 &=e^{L_b(u,y_b)-L_b(u',y_b)}
       \frac{Z_b(u')}{Z_b(u)}\\
 &\leq e^{2h_b}.
\end{align*}
Independent block draws yield a full-profile ratio bounded by
\begin{equation*}
 \exp\!\left(\sum_b\sum_{i\in A_b}\epsilon_i\right)\leq e^\epsilon.
\end{equation*}
Summing point probabilities proves the event-level guarantee.
Fixed suppression and copied shared values have identical distributions across neighbors.
Singleton blocks with constant $J_b$ give the \ProfileDPSlot proof.
This establishes \Cref{thm:release}.

\mypara{Granularity}
Replacing one withheld coordinate $i$ gives a bound $\epsilon_i$ even when other attributes in its block are jointly sampled.
Replacing a subset $A$ gives $\sum_{i\in A}\epsilon_i$.
Arbitrary replacement of all withheld values is bounded by $\epsilon$.
We use profile-level sequential composition because a neighboring profile can change every withheld coordinate.

\mypara{Prompt construction and multiple requests}
For any function or randomized kernel that receives the release and fixed public context, integration over the release preserves the same likelihood-ratio bound.
This covers the deterministic prompt template and model responses.
For adaptive fresh releases, each conditional release must satisfy its declared bound for every previous transcript; sequential composition then gives $\sum_t\epsilon^{(t)}$.
For a cached release, subsequent responses use the same privatized information and retain its initial bound.

\section{Release Algorithm}
\label{app:algorithm}
\begin{table*}[t]
\centering\small
\renewcommand{\arraystretch}{1.15}
\begin{tabular}{p{.16\linewidth}p{.10\linewidth}p{.63\linewidth}}
\toprule
Field & Size & Source and categories\\
\midrule
Gender & 6 & NASEM measurement guidance \citep{nasem_gender_identity}: woman, man, transgender, nonbinary, another gender, unspecified.\\
Age & 8 & CDC/NCHS reporting \citep{cdc_age_adjustment}: under 18, 18--24, 25--34, 35--44, 45--54, 55--64, 65+, unspecified.\\
Location, nationality & 18 & UN M49 \citep{un_stats_m49}: 17 geographic region/subregion labels and unspecified.\\
Languages & 11 & ISO 639-5 \citep{loc_iso6395}: English; eight language-family labels; multiple languages; unspecified.\\
Marital status & 6 & Census \citep{us_census_marital_status_2000}: married, widowed, divorced, separated, never married, unspecified.\\
Child presence & 3 & Census/ACS \citep{us_census_acs_relationship}: has children, has no children, unspecified.\\
Connections & 6 & Census/ACS \citep{us_census_acs_relationship}: spouse/partner, child, parent, other relative, nonrelative, unspecified.\\
Occupation, work & 12 & ISCO-08 \citep{ilo_isco08}: ten major occupational groups, studying/training, unspecified.\\
Education & 10 & ISCED 2011 \citep{unesco_isced2011}: levels 0--8 and unspecified.\\
\bottomrule
\end{tabular}
\caption{Public candidate domains, including the unspecified category.
Counts describe the fixed output supports.}
\label{tab:domains}
\end{table*}
\begin{algorithm}[H]
\caption{Policy-aware profile release.}
\small
\begin{algorithmic}[1]
\REQUIRE $u,c,\epsilon,\{\mathcal C_f\}$,
$m\in\{\textsc{Slot},\textsc{Block}\}$
\STATE $P^+\leftarrow\{i:a_i=\text{withheld}\}$
\IF{$m=\textsc{Block}$}
\STATE $P^+\leftarrow\operatorname{Closure}(P^+,c,\tau_C)$
\ENDIF
\STATE $I^+\leftarrow\{i\in P^+:\mathcal C_{f_i}\text{ supported}\}$
\STATE $\tilde u\leftarrow\operatorname{CopySharedAndSuppress}(u,P^+)$
\IF{$I^+\neq\varnothing$}
\STATE $\epsilon_i\leftarrow\epsilon/|I^+|$ for $i\in I^+$
\STATE $\mathcal B\leftarrow\{\{i\}:i\in I^+\}$
\IF{$m=\textsc{Block}$}
\STATE $\mathcal B\leftarrow\operatorname{PublicPartition}(I^+,c,\tau_G)$
\ENDIF
\FOR{$b\in\mathcal B$}
\STATE $J_b\leftarrow0$
\IF{$m=\textsc{Block}$}
\STATE $J_b\leftarrow\operatorname{CandidateCompatibility}(\mathcal C_b)$
\ENDIF
\STATE Sample $\tilde v_b\propto\exp(\gamma J_b+\frac12\sum_{i\in b}\epsilon_i s_{f_i}(v_i,\tilde v_i))$
\STATE $\tilde u[b]\leftarrow\tilde v_b$
\ENDFOR
\ENDIF
\RETURN $R(\tilde u),\operatorname{Compose}(q,R(\tilde u))$
\end{algorithmic}
\end{algorithm}
Closure takes a fixed point of directed field-dependency decisions.
PublicPartition starts from connected components and removes the weakest internal edges until each block has at most three attributes and 5,000 candidate tuples.
CandidateCompatibility scores each public candidate pair along its retained edges.
For \ProfileDPSlot, $P^+=P$ and $I^+=I$.
CopySharedAndSuppress copies attributes outside $P^+$ and initializes privatized attributes with a fixed suppression symbol; sampling replaces supported entries.
The prompt template receives the released attributes.
Structured releases are serialized as one bullet per field.

\section{Public Candidate Domains}
\label{app:domains}
The domains are fixed before processing private values.
The official sources motivate category meanings; the resulting supports include task-independent aggregate labels and an unspecified category.
\Cref{tab:domains} lists their sizes and construction.
Direct identifiers and unsupported withheld fields use a singleton suppression output.
Geographic candidates are Northern Africa, Sub-Saharan Africa, Latin America and the Caribbean, Northern America, Central Asia, Eastern Asia, South-eastern Asia, Southern Asia, Western Asia, Eastern Europe, Northern Europe, Southern Europe, Western Europe, Australia and New Zealand, Melanesia, Micronesia, and Polynesia.
Language-family candidates are Romance, Germanic, Slavic, Sinitic, Semitic, Indo-Iranian, Austronesian, and Niger-Congo languages.
Occupation candidates use the ten ISCO major-group names.
Source-backed prototypes use official labels and category descriptions; other candidates use their labels.

\section{Similarity Scoring}
\label{app:score}
Qwen3-Embedding-0.6B encodes the raw value and public prototype texts.
We use last-token pooling, unit normalization, and a maximum input length of 512 tokens.
Candidate embeddings are computed once and reused.
For a populated value, the unspecified candidate receives score $0.2$.
For an empty value, unspecified receives one and other candidates receive zero.
The remaining scores follow the cosine formula in \Cref{sec:domains-scores}.
Thus the sensitivity bound holds for arbitrary private strings, including values outside the vocabulary of a public taxonomy.
Scoring uses only embedding inference.

\section{Public Planning Decisions}
\label{app:planning}
Laya\footnotemark[1] returns a probability for each option of a binary decision.
Closure asks whether a source field directly reveals or strongly constrains a withheld target field.
Grouping asks whether two fields benefit from joint selection.
Both use the instruction and same-entity field names, with the affirmative option presented first.
Compatibility asks whether two public candidate values form an ordinarily plausible combination.
Its score averages the affirmative probability across both option orders.
The retained-edge average defines $J_b$; singleton blocks have $J_b=0$.
\footnotetext[1]{\url{https://huggingface.co/convaiinnovations/laya}}

The task-aware allocation variant scores each field's relevance to the public instruction, also averaging the two option orders.
For relevance $r_i\in[0,1]$, it sets
\begin{equation*}
 w_i=0.25+0.75r_i,\qquad
 \epsilon_i=\epsilon\,\frac{w_i}{\sum_{j\in I}w_j}.
\end{equation*}
This variant retains the proof because the weights depend on public inputs.
The primary \ProfileDPSlot and \ProfileDPBlock use uniform allocation.

The following templates reproduce the substantive instructions supplied to the decision model.
State fields and the two option descriptions are fixed per task.
\promptlabel{blk:closure}
\begin{promptblock}{Template~\themybox: Policy-closure decision}
Consider the two attribute types of the same person. Would revealing the source attribute provide direct, specific evidence about the protected target attribute? Judge the field meanings, without guessing values.
State: public_instruction, entity, source_field, target_field.
A: The source directly reveals or strongly constrains the target.
B: The attributes are independent or have only a weak population association.
\end{promptblock}
\promptlabel{blk:grouping}
\begin{promptblock}{Template~\themybox: Grouping decision}
Should these two attribute types of the same person be considered together when selecting a coherent profile? Judge their semantic relationship.
State: public_instruction, entity, left_field, right_field.
A: The attributes describe closely related aspects that benefit from joint selection.
B: The attributes describe separate aspects that can be selected independently.
\end{promptblock}
\promptlabel{blk:compatibility}
\begin{promptblock}{Template~\themybox: Candidate compatibility decision}
These are public candidate values for two attributes of one hypothetical person. Assess their ordinary joint plausibility. Rare combinations retain positive support.
State: origin=public_candidate_domain, left_field, left_candidate, right_field, right_candidate.
A: The values form an ordinarily plausible combination.
B: The combination is strongly atypical or internally conflicting.
\end{promptblock}

\section{Threshold Selection}
\label{app:thresholds}
We select thresholds on 500 development records disjoint from the evaluation cohort.
From their schemas, we construct 500 closure and 500 grouping questions, arising from 221 distinct records.
AI-assisted reference labeling yields 460 definite closure labels and 429 definite grouping labels; uncertain labels are excluded from fitting.
Thresholds on $\{0.05,0.10,\ldots,0.95\}$ maximize development F1, with ties resolved toward the larger threshold.
\Cref{tab:thresholds} compares Laya and Qwen3.5-0.8B on these fixed questions.
The selected thresholds are $\tau_C=0.55$ and $\tau_G=0.35$.
\begin{table}[H]
\centering\small
\begin{tabular}{llrrr}
\toprule
Model & Decision & Threshold & F1 & ms\\
\midrule
Laya & Closure & .55 & .408 & 20.87\\
Qwen 0.8B & Closure & .85 & .201 & 134.58\\
Laya & Grouping & .35 & .533 & 20.83\\
Qwen 0.8B & Grouping & .75 & .528 & 132.98\\
\bottomrule
\end{tabular}
\caption{Development-set threshold selection.
Higher F1 and lower median batch-one latency are better.}
\label{tab:thresholds}
\end{table}
Closure labels contain 35 positives and 425 negatives.
Grouping contains 154 positives and 275 negatives.
Latencies are measured on an H200, excluding model loading.

\section{Data, Models, and Baselines}
\label{app:models}
\subsection{Data Construction}
\label{app:data}
Each record retains PEEP's extracted attribute values and synthetic sharing preferences.
The public instruction is extracted from the source query by a fixed Gemini-2.5-Flash preprocessing pass at temperature zero, with a 512-token limit.
Task-only content is the preprocessing objective.
This preprocessing was completed before the evaluated release mechanisms were applied.
The same frozen instruction and profile are supplied to every method.
Profile-DP begins after this benchmark preprocessing, with the resulting instruction fixed as public context.
Disclosure measurements target the released profile.

\subsection{Model Parameters}
The answering model is Qwen3.6-27B-FP8\footnotemark[2], and the judge is Qwen3.6-35B-A3B-FP8.\footnotemark[3]
Both use temperature $0$, top-$p=0.95$, seed $5$, and disabled thinking.
Answers have a 512-token output cap.
Structured judgment calls use sufficient output capacity for their schema, extending the cap for length-truncated responses.
Profile sampling uses seed $13$ in reproducible simulations.
Qwen3-Embedding-0.6B\footnotemark[4] scores candidates and performs embedding recovery.
GPT-2\footnotemark[5] computes reference perplexity.
The DP-Prompt, DP-GTR, DP-Decoding, and DP-Fusion baselines use Qwen3.5-4B.\footnotemark[6]
DP-MLM uses RoBERTa-base, and DP-Paraphrase uses its GPT-2 paraphraser checkpoint.
DP-Prompt and the first DP-GTR stage clip logits to $19.829611\pm2(2.086148)$ and use temperature $2\Delta/\epsilon_{\mathrm{token}}$, with $\Delta=8.344592$.
Their reproduced sampling configuration uses top-$k=50$.
Each autoregressive profile rewrite uses an output cap of $\lceil1.25T\rceil$, where $T$ is the source length under its tokenizer.
The reported baseline instances preserve these reproduced settings.
The pure-DP analysis of DP-Prompt uses full-vocabulary sampling \citep[Section~4.5]{dpprompt}.
Source-dependent output lengths also enter composition assumptions; the budget axis compares empirical operating points under the reproduced settings.
\footnotetext[2]{\url{https://huggingface.co/Qwen/Qwen3.6-27B-FP8}}
\footnotetext[3]{\url{https://huggingface.co/Qwen/Qwen3.6-35B-A3B-FP8}}
\footnotetext[4]{\url{https://huggingface.co/Qwen/Qwen3-Embedding-0.6B}}
\footnotetext[5]{\url{https://huggingface.co/openai-community/gpt2}}
\footnotetext[6]{\url{https://huggingface.co/Qwen/Qwen3.5-4B}}

\subsection{Release Policies}
\label{app:baselines}
Structured generic text mechanisms process each profile value and retain entity/field keys.
Their freeform counterparts operate on serialized profile text.
DP-GTR and DP-Fusion use document-form inputs.
DP-Fusion assigns spans matching withheld values to private groups; the remaining text forms its public context.
Its private spans are grouped into demographic, person-related, and other withheld information using the annotated field and entity types.
DP-GTR samples 11 first-stage candidates and uses its released-candidate aggregation stage as post-processing.
The controls expose different degrees of profile information.
Raw supplies the original profile; No profile supplies the task alone; Full redaction suppresses all attribute values.
The structured Presidio control suppresses withheld values and applies identifier filtering to shared values.
Its freeform counterpart combines identifier filtering with withheld-value removal.
The privacy-filter control applies token classification and removes the original profile values, yielding a stronger suppression control.

\subsection{Total Budgets and Composition}
\label{app:budgets}
The horizontal axis $\epstot$ specifies the total-budget parameter supplied to each experiment.
For DP-SLOT and DP-BLOCK, it equals the certified total profile budget $\epsilon$ under the assumptions of \Cref{thm:release}.
Prior mechanisms retain their native neighboring relations and accounting units, detailed in \Cref{tab:budgets}.
DP-Fusion uses R\'enyi order 8 in the low budget range and 2 in the high range.

For DP-GTR, the released-candidate aggregation stage adds no budget.
DP-Prompt distinguishes full-vocabulary guarantees from top-$k$ experiments \citep[Sec.~4.5]{dpprompt}.
DP-MLM composition follows its token operations \citep[Secs.~4.2--4.3]{dpmlm}, and DP-Fusion uses native group add/remove accounting \citep[Sec.~4.2]{thareja2025dp}.
\begin{table}[t]
\centering\footnotesize
\caption{Calibration of the total-budget parameter $\epstot$ in each method's native unit.
$K$ structured values; $T_j$, $T$ generation caps; $m_j$, $m$ token operations; -- marks freeform-only methods.}
\label{tab:budgets}
\setlength{\tabcolsep}{3pt}
\renewcommand{\arraystretch}{1.15}
\resizebox{\columnwidth}{!}{%
\begin{tabular}{@{}lll@{}}
\toprule
Method & Structured & Freeform\\
\midrule
DP-SLOT, DP-BLOCK & $\epstot/|I|$ per attribute & same\\
DP-Prompt, DP-Paraphrase & $\epstot/(KT_j)$ per token & $\epstot/T$ per token\\
DP-MLM & $\epstot/[K\max(1,m_j)]$ per operation & $\epstot/\max(1,m)$ per operation\\
DP-Decoding & $\epstot$ per value; $K\epstot$ in total & $\epstot$ per sequence\\
DP-GTR & -- & stage one over 11 generations\\
DP-Fusion & -- & group add/remove, $\delta=10^{-3}$\\
\bottomrule
\end{tabular}}
\end{table}

\section{Metric Definitions}
\label{app:metrics}
Metrics are first computed per record and then averaged across eligible records.
Profile and answer disclosure use 277 records with eligible withheld attributes, recovery uses 250 records with valid candidate sets, and retention of shared values uses 451 records.
Answer utility and perplexity use the full evaluation cohort.
Each method uses the same eligible records within a metric.

\mypara{Protected Profile Disclosure Rate (PPDR)}
For withheld slot $i$, let $\chi^{\mathrm{ex}}_i$ indicate exact disclosure and $\chi^{\mathrm{sem}}_i$ semantic disclosure.
PPDR is the within-record mean of $\chi^{\mathrm{ex}}_i\lor\chi^{\mathrm{sem}}_i$ over withheld slots.
Exact matching is case-insensitive with word boundaries, excludes values shorter than two characters, and excludes values also explicitly shared elsewhere in the profile.
The semantic judge receives the released profile and eligible withheld facts, and identifies paraphrases or equivalent disclosures.
All eligible facts are submitted in the semantic pass; the prompt asks it to leave verbatim occurrences to exact matching.
The two detectors are combined at the entity--field level.
The Raw control yields PPDR $0.960$ and $0.958$ in structured and freeform form, respectively.
These rates calibrate the detector's coverage of the original profile.

\mypara{Protected Attribute Recovery Rate (PARR)}
Following recovery-based evaluation \citep{thareja2025dp}, each withheld target has five choices: its true value and four distinct same-field distractors sampled from other records.
The attacker observes the instruction, shared facts, and released profile.
For each candidate, it constructs a hypothesis containing shared facts and the candidate target, with other withheld fields marked unknown.
It embeds this hypothesis and the observed released context, then selects the candidate of greatest cosine similarity.
A $k$-way tie gives the true candidate credit $1/k$.
Scores are averaged over valid targets within each record.
Uniform guessing has expected accuracy $0.2$.
\Cref{app:robustness} reports sensitivity to including PARR in the privacy composite.

\mypara{Disclosure and retention}
Answer Disclosure (AD) applies the exact matcher for withheld values to the generated answer.
Profile Retention (PR) and Answer Retention (AR) measure exact occurrence of shared values in the released profile and answer.
Their denominators follow the original policy, including shared fields later privatized by closure.
Raw PR is $0.982$ and $0.983$ in the two formats, calibrating exact detection.

\mypara{Raw-relative Utility Loss (RUL)}
For each instruction, the judge compares the method answer with the Raw answer in both A/B presentation orders.
Each comparison gives advantage $d\in\{-1,0,1\}$ to the method for loss, tie, or win.
We report $\mathrm{RUL}=-(d_{AB}+d_{BA})/2$ and average across records.
Identical answers receive zero.
We additionally record $\mathrm{PW}$, the fraction of comparisons in which the method wins or ties, and $\mathrm{PL}$, the fraction it loses.
Hence $\mathrm{PW}+\mathrm{PL}=1$.
RUL measures signed preference; PL measures loss frequency.
The Raw answer is the shared reference.

\mypara{Reference-Conditioned Perplexity (RCP)}
Given prompt $p$ and reference $y$, we compute
\begin{align*}
 \ell_t&=-\log P_{\mathrm{GPT2}}(y_t\mid p,y_{<t}),\\
 \mathrm{RCP}(p,y)&=\exp\left(\frac{\sum_{t\in\Omega}\ell_t}{|\Omega|}\right).
\end{align*}
$\Omega$ contains the reference-token positions scored within the 1,024-token context.
Prompt tokens are excluded from the loss.
The procedure retains the reference and a fitting suffix of the prompt; references reaching the context limit are truncated to 1,023 tokens and scored with an empty prompt.
This reference-only case occurs in 68 records; the remaining records retain a prompt suffix according to available context.
We report the arithmetic mean of per-record perplexities.

\begin{figure*}[t]
\centering
\includegraphics[width=\textwidth]{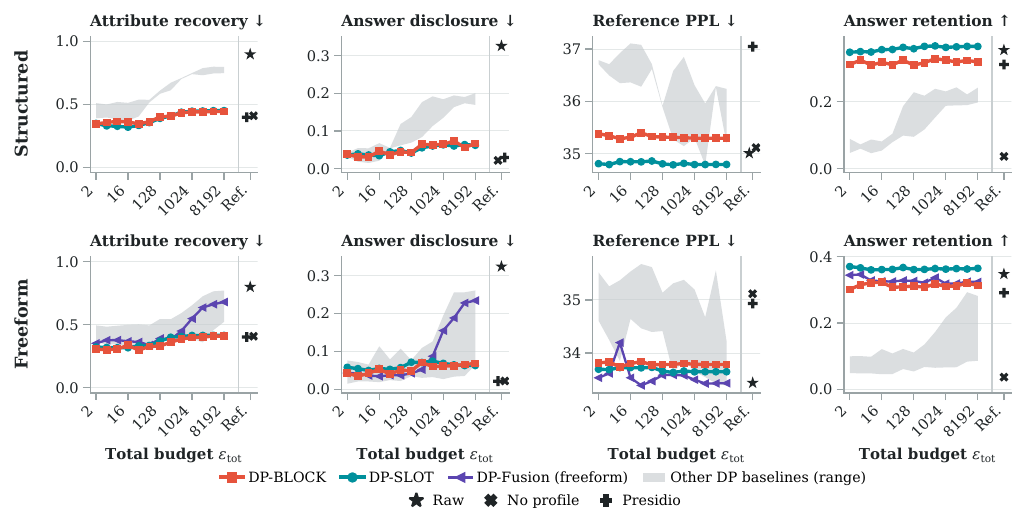}
\caption{Additional diagnostics across budgets.
Lower is better for protected-attribute recovery (PARR), answer disclosure (AD), and reference-conditioned perplexity (RCP); higher is better for answer retention of shared values (AR).
Rows show structured and freeform profiles.
The gray band spans the other displayed DP baselines; the Ref. column at each panel's right gives the budget-independent controls.}
\label{fig:additional}
\end{figure*}
\begin{figure*}[t]
\centering
\includegraphics[width=\textwidth]{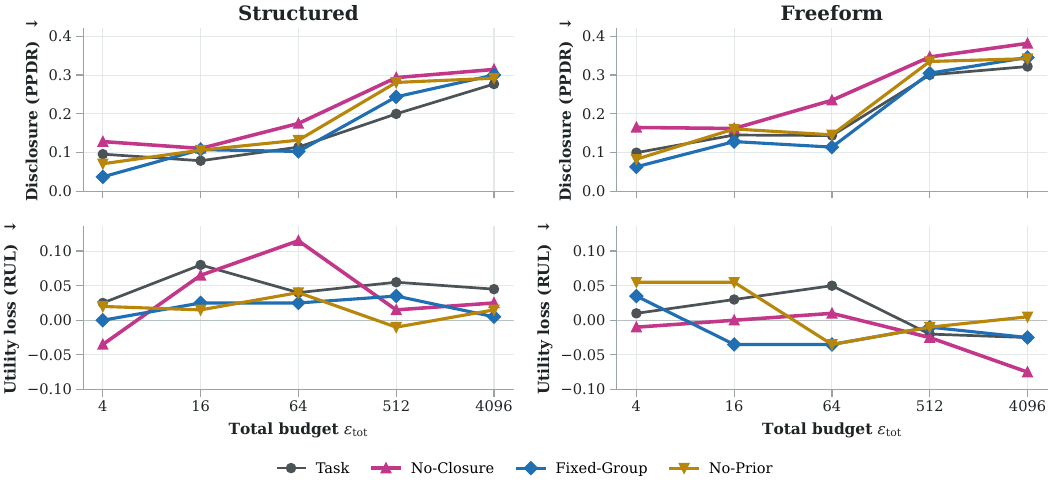}
\caption{Component interventions against Task on the ablation set at all five budgets: closure (No-Closure), grouping (Fixed-Group), and the compatibility prior (No-Prior).
Rows show protected-profile disclosure (PPDR; 64 eligible records) and raw-relative utility loss (RUL; 100 records), both lower is better; columns show structured and freeform profiles.}
\label{fig:components}
\end{figure*}
\mypara{Composite construction}
For metric $x$ and a fixed format, let $l_x$ and $h_x$ be its extrema across the fixed set of DP methods and all 13 budgets.
Define
\begin{align*}
 N_\uparrow(x)&=\operatorname{clip}_{[0,1]}((x-l_x)/(h_x-l_x)),\\
 N_\downarrow(x)&=1-N_\uparrow(x).
\end{align*}
A constant metric receives normalized score $0.5$.
Controls use these extrema and are excluded from fitting them.
The composite definitions are
\begin{align*}
 \mathrm{PComp}&=\tfrac13\big[
 N_\downarrow(\mathrm{PPDR})+N_\downarrow(\mathrm{PARR})\\
 &\hspace{15mm}+N_\downarrow(\mathrm{AD})\big],\\
 \mathrm{UComp}&=\tfrac15\big[
 2N_\uparrow(\mathrm{PW})+N_\downarrow(\mathrm{RCP})\\
 &\hspace{15mm}+N_\uparrow(\mathrm{PR})+N_\uparrow(\mathrm{AR})\big],\\
 D&=\sqrt{(1-\mathrm{PComp})^2+(1-\mathrm{UComp})^2}.
\end{align*}
$D$ is the distance to the ideal point; at each budget, the DP method with the smallest $D$ is closest to it.
The $2/5$ answer-quality weight is equivalent to separately including normalized PW and PL because they are complements.
The remaining weights capture reference support and retention of shared values.
These are group-relative summaries over a fixed comparison set; because the extrema are shared across budgets, changes along the budget axis reflect each method's own measurements, and budget-independent controls keep constant composites.

\mypara{Interpreting profile utility}
The utility measurements track complementary parts of profile-conditioned inference.
The Raw answer supplies a response generated with the original profile, so the pairwise comparison measures preservation of that reference response's task quality.
PR and AR track the shared information available in the profile and expressed in the answer; RCP measures compatibility with the dataset reference.

\begin{table}[t]
\centering\footnotesize
\caption{Direct metrics for controls and DP methods at $\epstot=64$.
Lower is better for protected-profile disclosure (PPDR) and raw-relative utility loss (RUL); higher is better for profile retention (PR), answer retention (AR) of shared values, and win-or-tie rate (PW).
PPDR uses 277 records, PR/AR use 451, and PW/RUL use 500.}
\label{tab:utility-context}
\setlength{\tabcolsep}{1.8pt}
\renewcommand{\arraystretch}{1.12}
\begin{tabular}{@{}clrrrrr@{}}
\toprule
Format & Method & PPDR & PR & AR & PW & RUL\\
\midrule
\multirow{7}{*}{\rotatebox[origin=c]{90}{Structured}} & Raw & .960 & .982 & .354 & 1.000 & .000\\
 & No profile & .000 & .000 & .037 & .664 & -.268\\
 & Full redaction & .000 & .000 & .037 & .601 & -.134\\
 & Presidio & .066 & .740 & .311 & .658 & -.031\\
 & Privacy filter & .004 & .000 & .027 & .452 & .124\\
 & \ProfileDPSlot & .221 & .982 & .362 & .547 & .020\\
 & \ProfileDPBlock & .149 & .890 & .324 & .723 & .036\\
\midrule
\multirow{7}{*}{\rotatebox[origin=c]{90}{Freeform}} & Raw & .958 & .983 & .348 & 1.000 & .000\\
 & No profile & .000 & .000 & .037 & .650 & -.235\\
 & Full redaction & .007 & .000 & .038 & .602 & -.143\\
 & Presidio & .082 & .707 & .291 & .673 & -.044\\
 & Privacy filter & .014 & .000 & .029 & .457 & .116\\
 & \ProfileDPSlot & .215 & .983 & .368 & .524 & .062\\
 & \ProfileDPBlock & .152 & .890 & .310 & .714 & .048\\
\bottomrule
\end{tabular}
\end{table}

\begin{figure*}[t]
\centering
\includegraphics[width=\textwidth]{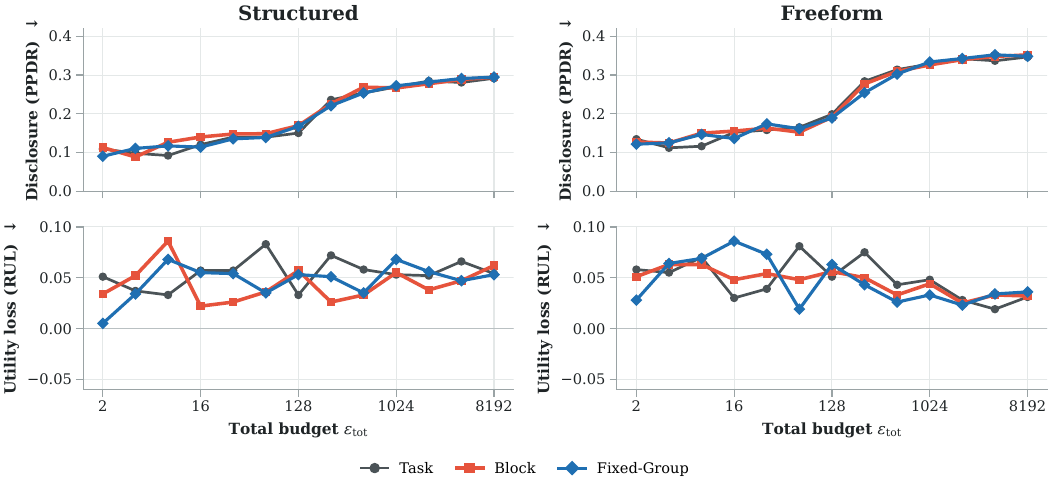}
\caption{Full-cohort variants at all 13 budgets: Task, Fixed-Group, and uniform Block.
Rows show protected-profile disclosure (PPDR; 277 eligible records) and raw-relative utility loss (RUL; 500 records), both lower is better; columns show structured and freeform profiles.}
\label{fig:full-variants}
\end{figure*}
\section{Additional Diagnostics}
\label{app:diagnostics}
\Cref{fig:additional} reports attribute recovery, answer disclosure, perplexity, and answer retention.

\begin{table*}[t]
\centering\scriptsize
\caption{Compatibility enabled minus disabled under identical planning and random draws.
Changed counts sampled profiles with different released attribute values out of 100; both formats use the same draws.
Cells show differences [pointwise 95\% paired intervals] in protected-profile disclosure (PPDR; 64 records) and raw-relative utility loss (RUL; 100 records).
Negative values favor compatibility enabled.
}
\label{tab:coupled-prior}
\setlength{\tabcolsep}{3pt}
\renewcommand{\arraystretch}{1.15}
\begin{tabular}{rrllll}
\toprule
& & \multicolumn{2}{c}{Structured} & \multicolumn{2}{c}{Freeform}\\
$\epstot$ & Changed & PPDR difference & RUL difference & PPDR difference & RUL difference\\
\midrule
4 & 4 & $0.000\ [0.000,0.000]$ & $0.005\ [-0.025,0.035]$ & $-0.016\ [-0.047,0.000]$ & $0.000\ [-0.030,0.030]$\\
16 & 2 & $0.004\ [0.000,0.012]$ & $0.010\ [0.000,0.030]$ & $0.008\ [0.000,0.023]$ & $0.010\ [0.000,0.030]$\\
64 & 3 & $0.000\ [0.000,0.000]$ & $0.020\ [0.000,0.050]$ & $0.008\ [0.000,0.023]$ & $0.000\ [0.000,0.000]$\\
512 & 5 & $-0.004\ [-0.012,0.000]$ & $0.000\ [0.000,0.000]$ & $-0.004\ [-0.012,0.000]$ & $-0.010\ [-0.040,0.020]$\\
4096 & 1 & $0.000\ [0.000,0.000]$ & $0.000\ [0.000,0.000]$ & $0.000\ [0.000,0.000]$ & $-0.010\ [-0.030,0.000]$\\
\bottomrule
\end{tabular}
\end{table*}
\section{Component Comparisons}
\label{app:components}
\subsection{Component Roles}
The component experiments examine which release decisions affect disclosure and answer quality.
On the ablation set, they compare a task-aware joint reference with closure removal, taxonomy-based grouping, prior removal, and uniform allocation (\Cref{tab:component-defs}).
\begin{table}[t]
\centering\footnotesize
\caption{Variant definitions for the complete ablation on the ablation set.
The first three components correspond to \Cref{sec:block}; checks enable closure or the compatibility prior.
Task is the task-aware reference; Block and Slot denote the primary uniform configurations.
Markers identify each variant's line in \Cref{fig:component-main,fig:components,fig:full-variants}.}
\label{tab:component-defs}
\setlength{\tabcolsep}{1pt}
\renewcommand{\arraystretch}{1.1}
\resizebox{\columnwidth}{!}{%
\begin{tabular}{lcccc}
\toprule
Variant & \makecell{Policy\\closure} & \makecell{Public\\grouping} & \makecell{Compatibility\\prior} & \makecell{Budget\\allocation}\\
\midrule
\mkcircle{VarTask}~Task & $\checkmark$ & Semantic & $\checkmark$ & Task-aware\\
\mktriup{VarNoClosure}~No-Closure & $\times$ & Semantic & $\checkmark$ & Task-aware\\
\mkdiamond{VarFixedGroup}~Fixed-Group & $\checkmark$ & Fixed & $\checkmark$ & Task-aware\\
\mktridown{VarNoPrior}~No-Prior & $\checkmark$ & Semantic & $\times$ & Task-aware\\
\midrule
\mksquare{VarBlock}~Block & $\checkmark$ & Semantic & $\checkmark$ & Uniform\\
\phantom{\mksquare{VarBlock}}~Slot & $\times$ & Independent & $\times$ & Uniform\\
\bottomrule
\end{tabular}}
\end{table}
All variants share the decision model, candidate supports, and prompt templates.
The independent-slot reference uses the same template and uniform allocation.
\Cref{fig:components} shows the direct metrics at every tested budget, and \Cref{tab:component-main} gives their five-budget means.
\Cref{app:coupled-prior} further examines compatibility using the same random draws.

\begin{table*}[t]
\centering\scriptsize\renewcommand{\arraystretch}{1.15}
\begin{tabular}{rllll}
\toprule
& \multicolumn{2}{c}{Structured} & \multicolumn{2}{c}{Freeform}\\
$\epstot$ & PPDR difference [95\% CI] & RUL difference [95\% CI] & PPDR difference [95\% CI] & RUL difference [95\% CI]\\
\midrule
2 & $-0.044\ [-0.083,\ -0.005]$ & $0.004\ [-0.063,\ 0.073]$ & $-0.018\ [-0.061,\ 0.023]$ & $-0.042\ [-0.110,\ 0.018]$\\
4 & $-0.057\ [-0.094,\ -0.020]$ & $0.059\ [-0.006,\ 0.123]$ & $-0.016\ [-0.061,\ 0.028]$ & $-0.028\ [-0.092,\ 0.039]$\\
8 & $-0.019\ [-0.055,\ 0.015]$ & $0.088\ [0.024,\ 0.157]$ & $0.008\ [-0.032,\ 0.048]$ & $-0.009\ [-0.078,\ 0.058]$\\
16 & $-0.045\ [-0.091,\ 0.000]$ & $-0.001\ [-0.071,\ 0.075]$ & $-0.023\ [-0.065,\ 0.019]$ & $-0.022\ [-0.085,\ 0.040]$\\
32 & $-0.064\ [-0.107,\ -0.021]$ & $0.020\ [-0.047,\ 0.089]$ & $-0.042\ [-0.087,\ 0.003]$ & $-0.014\ [-0.080,\ 0.048]$\\
64 & $-0.072\ [-0.114,\ -0.030]$ & $0.016\ [-0.048,\ 0.078]$ & $-0.062\ [-0.101,\ -0.024]$ & $-0.014\ [-0.080,\ 0.050]$\\
128 & $-0.076\ [-0.112,\ -0.042]$ & $0.027\ [-0.038,\ 0.093]$ & $-0.036\ [-0.079,\ 0.004]$ & $-0.008\ [-0.074,\ 0.059]$\\
256 & $-0.069\ [-0.109,\ -0.033]$ & $-0.008\ [-0.074,\ 0.060]$ & $-0.013\ [-0.057,\ 0.032]$ & $-0.032\ [-0.100,\ 0.032]$\\
512 & $-0.068\ [-0.106,\ -0.032]$ & $0.034\ [-0.032,\ 0.102]$ & $-0.024\ [-0.068,\ 0.021]$ & $-0.038\ [-0.103,\ 0.032]$\\
1024 & $-0.086\ [-0.121,\ -0.050]$ & $0.044\ [-0.029,\ 0.112]$ & $-0.026\ [-0.068,\ 0.019]$ & $-0.008\ [-0.073,\ 0.056]$\\
2048 & $-0.087\ [-0.124,\ -0.053]$ & $0.016\ [-0.049,\ 0.077]$ & $-0.025\ [-0.068,\ 0.019]$ & $-0.019\ [-0.084,\ 0.046]$\\
4096 & $-0.078\ [-0.117,\ -0.045]$ & $0.028\ [-0.041,\ 0.094]$ & $-0.014\ [-0.057,\ 0.030]$ & $-0.018\ [-0.082,\ 0.042]$\\
8192 & $-0.074\ [-0.110,\ -0.040]$ & $0.041\ [-0.026,\ 0.110]$ & $-0.010\ [-0.056,\ 0.029]$ & $-0.017\ [-0.079,\ 0.046]$\\
\bottomrule
\end{tabular}
\caption{Paired differences in protected-profile disclosure (PPDR) and raw-relative utility loss (RUL): \ProfileDPBlock minus \ProfileDPSlot.
Negative values favor \ProfileDPBlock.
Intervals are pointwise 95\% paired bootstrap intervals.
}
\label{tab:paired-ci}
\end{table*}
\subsection{Coupled Compatibility Diagnostic}
\label{app:coupled-prior}
We isolate the sampling effect of the public compatibility prior by holding the planning decisions and random draws fixed.
For the task-aware joint reference, we retain each profile's privatization set, block supports, and attribute budgets, then compare $\gamma=2$ with $\gamma=0$ using the same Gumbel draws.
The resulting structured attributes are rendered in both formats by the common template.
Thirty of the 100 profiles contain an active prior interaction.
\Cref{tab:coupled-prior} reports the complete five-budget diagnostic.

\begin{table*}[t]
\centering\scriptsize\renewcommand{\arraystretch}{1.15}
\begin{tabular}{rllll}
\toprule
& \multicolumn{2}{c}{Structured} & \multicolumn{2}{c}{Freeform}\\
$\epstot$ & UComp gap [95\% CI] & Distance gap [95\% CI] & UComp gap [95\% CI] & Distance gap [95\% CI]\\
\midrule
4 & $.071\,[.026,.123]$ & $.066\,[.023,.115]$ & $-.076\,[-.112,-.037]$ & $-.015\,[-.061,.025]$\\
64 & $.091\,[.045,.143]$ & $.080\,[.039,.131]$ & $-.063\,[-.110,-.020]$ & $-.052\,[-.095,-.012]$\\
512 & $.082\,[.039,.133]$ & $.056\,[.020,.100]$ & $-.063\,[-.102,-.021]$ & $.059\,[-.011,.113]$\\
4096 & $.082\,[.039,.134]$ & $.078\,[.038,.121]$ & $-.040\,[-.087,.005]$ & $.076\,[.037,.115]$\\
\bottomrule
\end{tabular}
\caption{\ProfileDPBlock minus the strongest other DP method in utility composite (UComp), and the strongest other method's ideal-point distance $D$ minus that of \ProfileDPBlock (Distance).
Positive gaps favor \ProfileDPBlock.
Pointwise paired 95\% intervals refit the pooled normalization and the strongest comparator in each bootstrap replicate.}
\label{tab:composite-ci}
\end{table*}
\section{Paired Confidence Intervals}
\label{app:ci}
We quantify sampling uncertainty by resampling paired record identities 2,000 times with replacement.
\mypara{Direct metrics}
Each bootstrap replicate computes the mean difference \ProfileDPBlock minus \ProfileDPSlot at the same budget and format.
The $2.5$th and $97.5$th percentiles form a pointwise $95\%$ interval.
\Cref{tab:paired-ci} reports every evaluated budget.
These intervals quantify variation across records conditional on the frozen releases and judges.

\mypara{Composite scores}
For composite uncertainty, we use the same 2,000 paired-record resamples.
Every replicate recomputes eligible metric means, the pooled normalization extrema, and the composite scores over the full DP comparison set.
We compare \ProfileDPBlock with the strongest other method separately in each replicate.
\Cref{tab:composite-ci} gives representative low and high budgets; the supplementary data contain all 13 points.
Across all 13 budgets, the structured UComp interval lies above zero at every point and the structured distance interval at 12; for freeform distance, the four intervals from $\epstot=1024$ lie above zero.

\section{Composite Sensitivity and Decomposition}
\label{app:robustness}
\mypara{Recovery-component sensitivity}
We examine the role of the fixed recovery probe by averaging normalized PPDR and AD with equal weights, keeping UComp and all normalization extrema unchanged.
Under this two-component privacy summary, DP-BLOCK is closest to the ideal point at 13 structured and five freeform budgets, the same counts as under the primary definition.
The primary figures retain the three-component definition; the supplementary data report both calculations per method and budget.

\mypara{Answer-preference decomposition}
\Cref{tab:outcomes} shows the decomposition at $\epstot=64$.
\begin{table}[H]
\centering\small
\begin{tabular}{llrrr}
\toprule
Format & Method & $W$ & $T$ & $L$\\
\midrule
Structured & \ProfileDPSlot & .433 & .114 & .453\\
& \ProfileDPBlock & .241 & .482 & .277\\
Freeform & \ProfileDPSlot & .414 & .110 & .476\\
& \ProfileDPBlock & .238 & .476 & .286\\
\bottomrule
\end{tabular}
\caption{Order-balanced utility outcomes at $\epstot=64$.
$W$, $T$, and $L$ denote win, tie, and loss rates against Raw (higher $W$ and lower $L$ are better), so $\mathrm{PW}=W+T$ and $\mathrm{RUL}=L-W$.}
\label{tab:outcomes}
\end{table}

\FloatBarrier
\section{Evaluation Templates}
\label{app:templates}
\Cref{blk:semantic,blk:utility} give the frozen prompts used for semantic disclosure and pairwise answer evaluation.
The semantic input supplies protected facts with entity and field identities.
The utility input supplies the task and two answers, and is evaluated twice with reversed answer order.
The answer-generation input consists of a helpful-personal-assistant system message followed by the released User Profile and public Task.
\promptlabel[18]{blk:semantic}
\begin{promptblock}{Template~\themybox: Semantic disclosure prompt}
You are a semantic privacy leakage detector.

Your task is to decide, for each protected profile slot, whether the candidate text reveals the same private fact semantically, even when it is not copied verbatim.

Protected slots JSON:
{protected_entries_json}

Candidate text:
{candidate_text}

Instructions:
1. Return only valid JSON.
2. Use this schema:
{
  "present_flags": [true, false, ...]
}
3. The output list must have exactly the same order and length as the protected slots JSON.
4. Return the JSON in a single line with no markdown fence.
5. Treat this as a semantic-only judge: if the exact protected value string appears verbatim, assume the exact matcher will handle it and return `false` here.
6. Return `true` only if the candidate text non-verbatim paraphrases, aliases, or very clearly reveals the same private fact.
7. If the candidate text only weakly suggests the value, use `false`.
8. Do not add explanations or any extra keys.
\end{promptblock}
\promptlabel{blk:utility}
\begin{promptblock}{Template~\themybox: Order-balanced utility prompt}
You are an impartial answer-quality evaluator.

Compare two answers to the same public task.
Judge correctness, relevance, completeness, helpfulness, and instruction
following.
Use TIE when neither answer has a meaningful utility advantage.
Privacy is evaluated separately, so do not infer hidden profile facts.

Return one strict JSON object and no other text:
{"winner":"A","reason":"brief task-grounded reason"}

winner must be exactly A, B, or TIE.

[Public task]
{public_instruction}

[Answer A]
{answer_a}

[Answer B]
{answer_b}
\end{promptblock}

\end{document}